\documentclass[12pt,preprint]{aastex}

\newcommand{\gcc}{\mathrm{g\ cm^{-3} }}
\newcommand{\cms}{\mathrm{cm\ s^{-1}}}

\slugcomment{submitted to ApJ}

\newcommand{\Ma}{\mathrm{Ma}}
\usepackage{epsf,color,dcolumn}

\newcolumntype{d}{D{.}{.}{-1}}

\epsfverbosetrue

\begin{document} 

\title{Direct Numerical Simulations of Type Ia Supernovae Flames I: The
Landau-Darrieus Instability}

\shorttitle{LD Instability in Thermonuclear Flames}
\shortauthors{Bell et al.}

\author{J.~B.~Bell\altaffilmark{1},
        M.~S.~Day\altaffilmark{1},
        C.~A.~Rendleman\altaffilmark{1},
        S.~E.~Woosley\altaffilmark{2},
	M.~Zingale\altaffilmark{2}}

\altaffiltext{1}{Center for Computational Science and Engineering,
                 Lawrence Berkeley National Laboratory,
                 Berkeley, CA 94720}

\altaffiltext{2}{Dept.\ of Astronomy \& Astrophysics,
                 The University of California, Santa Cruz,
                 Santa Cruz, CA 95064}

\begin{abstract}

Planar flames are intrinsically unstable in open domains due to the
thermal expansion across the burning front---the Landau-Darrieus
instability.  This instability leads to wrinkling and growth of the
flame surface, and corresponding acceleration of the flame, until it
is stabilized by cusp formation.  We look at the Landau-Darrieus
instability for C/O thermonuclear flames at conditions relevant to the
late stages of a Type Ia supernova explosion.  Two-dimensional direct
numerical simulations of both single-mode and multi-mode perturbations
using a low Mach number hydrodynamics code are presented.  We show the
effect of the instability on the flame speed as a function of both the
density and domain size, demonstrate the existence of the small scale
cutoff to the growth of the instability, and look for the proposed
breakdown of the non-linear stabilization at low densities.  The
effects of curvature on the flame as quantified through measurements
of the growth rate and computation of the corresponding Markstein
number.  While accelerations of a few percent are observed, they are
too small to have any direct outcome on the supernova explosion.

\end{abstract}


% ============================================================================
%  Introduction
% ============================================================================
\section{INTRODUCTION}
\label{sec:intro}

A carbon/oxygen flame propagating outward from the center of a white
dwarf is subjected to a number of instabilities that wrinkle and
accelerate the flame.  If the flame can accelerate to a significant
fraction of the speed of sound, a deflagration alone can account for
the explosion (see \citealt{hillebrandtniemeyer2000} and references
therein; \citealt{reinecke2002b, gamezo2003}).  One such instability,
the Landau-Darrieus (LD) instability
\citep{darrieus:1938,landau:1944} affects a planar flame front, even
in the absence of gravity, driven by the thermal expansion across the
flame.  The LD unstable flame will wrinkle, eventually forming cusps
in the nonlinear regime that stabilize the flame.  The growth rate
for a LD unstable flame, including the effects of the finite
thickness of the flame, $l_f$, is
\begin{equation}
\label{eq:ld}
\omega = k U_l \frac{\alpha}{\alpha + 1} 
         \left [ \sqrt{\alpha - \frac{1}{\alpha} + 1 + k \; l_f \Ma (k
 \;l_f \Ma + 2 \alpha)} - 1 + k \; l_f \Ma \right ] 
\end{equation}
\citep{zeldovich85}, where $\alpha =
\rho_{\mathrm{fuel}}/\rho_{\mathrm{ash}}$ is the density ratio across
the flame, $k$ is the wavenumber, $U_l$ is the laminar flame speed,
and $\Ma$ is the Markstein number.  The Markstein number is a measure
of the response of the flame speed, $U$, to curvature,
\begin{equation}
U = U_l \left ( 1 + \Ma \; l_f \frac{\partial^2 y_f}{\partial x^2} \right ) \enskip ,
\end{equation}
where $y_f$ is the position of the flame interface and $x$ is the
transverse coordinate.  We note that our sign convention for $\Ma$ is
opposite that used in \citet{zeldovich85} (see
\citealt{flame-curvature} for a discussion of the Markstein number for
astrophysical flames).  This growth rate has been confirmed
experimentally \citep{clanet1998} for chemical flames.  In this paper,
we present two-dimensional direct numerical simulations (DNS) of the
LD instability for low-density C/O flames in white dwarf interiors.
We start by briefly reviewing previous LD simulations and results
relevant for thermonuclear flames.

\citet{sivashinsky1977} and \citet{michelson1977} studied the
non-linear regime of the LD instability analytically and numerically,
using an integro-differential equation to describe the deformation of
a discontinuous flame front.  This was extended later by
\citet{gutman1990}, who concentrated on the regime where thermal
conduction dominates over mass diffusion (this is the regime
appropriate for astrophysical flames).  They experiment with boundary
conditions (both adiabatic and periodic) and discuss the influence of
the domain size on the final cusp behavior.  Their simulations show a
multi-mode perturbation cusping, and slowly, the cusps merging until
only a single one survives.  In the widest domains, a superimposed
cellular structure appears on the single remaining cusp.  The
mechanism behind this cellular instability does not, at this moment,
seem to be well understood (although see \citealt{kupervasser}).  A
similar approach, using a more general equation for the front, the
Frankel equation, considering spherically expanding flames, was
performed by \citet{blinnikovsasorov}.  They observe a fractal
structure for the front through a range of spatial scales.  The
unstable flame exhibits cell-splitting, owing to the expanding
spherical geometry.  \citet{helenbrooklaw1999} presented simulations that
treat the fuel and ash states as incompressible, and link them with
jump conditions across the discontinuous front.  In contrast to other
methods, feedback between the flame and fluid flow is included in this
model.  They found that the flow dominates the LD instability in
setting the scale for the wrinkling.  This has important consequences
for the flame in Type Ia supernova (SNe Ia), as the medium will be turbulent,
and some evidence suggests that a large-scale dipole flow is setup
during the ignition phase of the explosion \citep{kuhlen2003}.  In the
real star, it may be likely that the fluid motions swamp any effects
of the LD instability.  Furthermore, the LD instability will be
competing with the Rayleigh-Taylor instability \citep{SNrt}.

Recently, front tracking methods treating a discontinuous flame
coupled to full hydrodynamics have been used for both terrestrial
\citep{qian1998} and astrophysical flames \citep{rhn2003} to
investigate LD unstable flames.  The work of \citet{rhn2003} used a
level set method coupled to fully compressible PPM \citep{ppm} to
follow a discontinuous flame front, using flame speeds from
\citet{timmeswoosley1992}.  In their highest resolution run, they
report seeing a cellular structure distorting the single cusp, perhaps
arising through the same process as described in \citet{gutman1990}.
\citet{rhn2003} find the flame speed rising to $1.3\times$ the laminar
speed.

The work described above all treat the flame as a discontinuous
front---some model is required to describe the behavior of the flame
on scales smaller than the grid resolution.  The finite thickness of
the flame is known to have some influence on the LD instability, such
as setting the small scale cutoff for the growth, so simulations that
resolve the thermal structure are important.  Direct numerical
simulations of the LD instability at various densities in the
astrophysical context were performed by
\citet{niemeyerhillebrandt1995a}, showing cusp formation, but it was
unclear from their studies whether there was any acceleration over the
laminar speed.  Their lowest density flame ($5\times 10^7~\gcc$)
showed significant growth beyond simple cusping, but it is likely that
this was either numerical in origin, or due to the interaction of the
transient perturbations left over from the initialization with the
flame.

LD unstable flames in laminar flows are unlikely alone to lead to the
accelerations required of the SNe Ia models.  An interesting question
is whether the non-linear stabilization can break down, whether
through some property intrinsic to the flames or through external
forces. \citet{kerstein1996} discusses one such mechanism, active
turbulent combustion, involving the interaction of the LD unstable
flame with turbulence.  It has been proposed that this process could
occur in Type Ia supernovae \citep{niemeyerwoosley1997}, since large
amounts of the thermonuclear energy released goes into the expansion
of the star, and it is this thermal expansion that drives the LD
instability.  \citet{niemeyerwoosley1997} argue that this feedback
between the flame and its own turbulence could lead to large accelerations of
the flame, although, at present, this has not been demonstrated in
numerical simulations.

We focus on DNS of the Landau-Darrieus instability at low densities
($\rho \sim 2\times 10^7~\gcc$ -- $8\times 10^7~\gcc$) using a low
Mach number numerical formulation, as described in
\citet{SNeCodePaper}.  Here we are able to extend to lower densities
than previous studies, and also, through the use of adaptive mesh
refinement, use finer resolution and larger domains.  For the present
study, we restrict ourselves to two-dimensional calculations.  These
direct numerical simulations are truly parameter free (i.e., no
laminar flame speed is input), and could serve to test\slash calibrate
different methods of representing unresolved flames on stellar-sized
domains.  Corresponding DNS calculations of the Rayleigh-Taylor
instability will be presented in a second paper \citep{SNrt}.
Three-dimensional results will be presented in a future paper.
Additionally, we do not look to seed the flow with turbulence in the
present study, but rather choose to focus solely on the pure LD
instability.  We provide a brief description of the numerics and
parameters used for the present simulations in \S\ref{sec:numerics}.
In \S\ref{sec:results}, we discuss the results, and finally, in
\S\ref{sec:conclusions} we conclude.

% ============================================================================
%  Numerical Methods
% ============================================================================
\section{NUMERICAL METHODS}
\label{sec:numerics}

The evolution was carried out using a low Mach number formulation of
the Euler equations, as described in \citet{SNeCodePaper,DayBell00}.
The low Mach number formulation takes advantage of the fact that for
slow moving flames, the pressure is constant across the flame front,
to a high degree.  Scaling the state variables and expanding them in
powers of Mach number, keeping terms to $O(M^2)$ \citep{MajSet85}
yields a system of equations whose numerical stability condition is
set by the fluid velocity alone, instead of the flow velocity plus
sound speed.  The pressure is decomposed into a dynamic and
thermodynamic component, the ratio of which is of $O(M^2)$, with only
the dynamic component appearing in the momentum equation.  The
equation of state is enforced by requiring the pressure to be constant
along particle paths, resulting in an elliptic constraint on the
velocity field, completing the system of equations.  For the flames we
consider in the present paper, $M < 10^{-4}$, so this approximation is
valid.  This allows for timesteps that are $O(1/M$), larger than a
corresponding compressible code.  

The low Mach number equations are solved via a second-order,
fractional-step advection/projection/reaction procedure.  The
advection step advances the fluid variables to the new timestep and
finds a provisional velocity field that does not yet satisfy the
elliptic constraint.  The new velocities are found by projecting this
velocity field onto the space that discretely satisfies the elliptic
constraint.  Reactions are included via Strang-splitting.  We refer
the reader to \citet{SNeCodePaper} for the full details of the
numerical method, as it is unchanged for the present study.  Laminar
astrophysical flames computed with this code were compared to the
benchmark calculations presented in \citet{flame-curvature}, and found
to be in good agreement.

The calculations we present here differ from the fully compressible
PPM \citep{ppm} method used in \citet{niemeyerhillebrandt1995a} in several
fundamental ways.  First, the low Mach number formulation yields a
timestep constraint that depends only on the fluid velocity, not the
fluid velocity and sound speed as in the compressible case.  This
means that we can take much larger timesteps, which for very slow
moving flames, greatly reduces the accumulation of error.
Additionally, the advection algorithm we employ is not dimensionally
split, reducing the influence of the grid on the flame.  Finally, the
two dimensional domain is initialized by mapping a steady state flame
in a perturbed fashion onto the grid, rather than perturbing an
interface between fuel and ash and waiting for the flame to develop.
This significantly reduces the transients and leaves the flow ahead of
the flame unperturbed.

An equation of state with degenerate\slash relativistic electrons and
positrons, ideal gas ions, and Boltzmann radiation was used, as
described in \citet{timmesswesty2000}.  The thermal conductivity is
described in \citet{timmes_he_flames:2000}.  A single
reaction,$^{12}$C($^{12}$C,$\gamma$)$^{24}$Mg, is followed, using the
unscreened rate from \citet{caughlan-fowler:1988}.  This accounts for
the bulk of the energy generation in a C/O flame.  Furthermore, at the
densities we are considering, burning is expected to terminate at
intermediate mass elements, so stopping at $^{24}$Mg is reasonable.
For this simple burning rate, we ignore the effects of the oxygen on
the energy generation---it is only advected passively here.  The
temperatures reached behind the flame at these densities are $\sim
4\times 10^9$~K.  At these temperatures, oxygen burning is several
orders of magnitude slower than carbon, and we would not expect any
effects from it on the timescales that we follow the flames.  The
properties of the flames we consider here are listed in
Table~\ref{table:flameproperties}.  We will consider the $4\times
10^7~\gcc$ flame to be our reference flame, upon which we will explore
the effects of resolution and domain size.

Motivated by results indicating that multi-mode perturbations
eventually merge into a single dominant cusp (see for example
\citealt{gutman1990,SNeCodePaper}), we concentrate mainly on single
mode perturbations.  In all cases, the flame is initialized on the
grid by mapping a steady-state one-dimensional laminar flame across
the grid, shifting it vertically according to the perturbation.  The
boundary conditions in all cases are periodic transverse to the flame
propagation direction, inflow ahead of the flame, and outflow behind
the flame.  The flame is oriented such that it is propagating downward
in our domain.  The inflow conditions correspond to the fuel
conditions, with the inflow velocity chosen to match the laminar flame
velocity.  Thus, a flame moving at the laminar speed will remain
stationary in the computational domain.  Any acceleration resulting
from the flame instability will lead to motion across the grid.  The
resolution used is chosen to put about 5--10 zones inside the thermal
width, $l_f$, calculated as
\begin{equation}
l_f = \frac{T_{\mathrm{ash}} - T_{\mathrm{fuel}}}{\mathrm{max}(\nabla T)} \enskip ,
\end{equation}
where $T_{\mathrm{ash}}$ and $T_{\mathrm{fuel}}$ are the ash and fuel
temperatures respectively.  We note that this flame width measure is
$\sim 2$~times narrower than that used in \citet{timmeswoosley1992}
(see also \citealt{flame-curvature}).  In \S \ref{sec:results}, we
look at the sensitivity of the results to the resolution.

% ============================================================================
%  Results 
% ============================================================================
\section{RESULTS}
\label{sec:results}

\subsection{Multi-mode perturbations}

\begin{figure}
\begin{center}
\epsscale{0.9}
\plotone{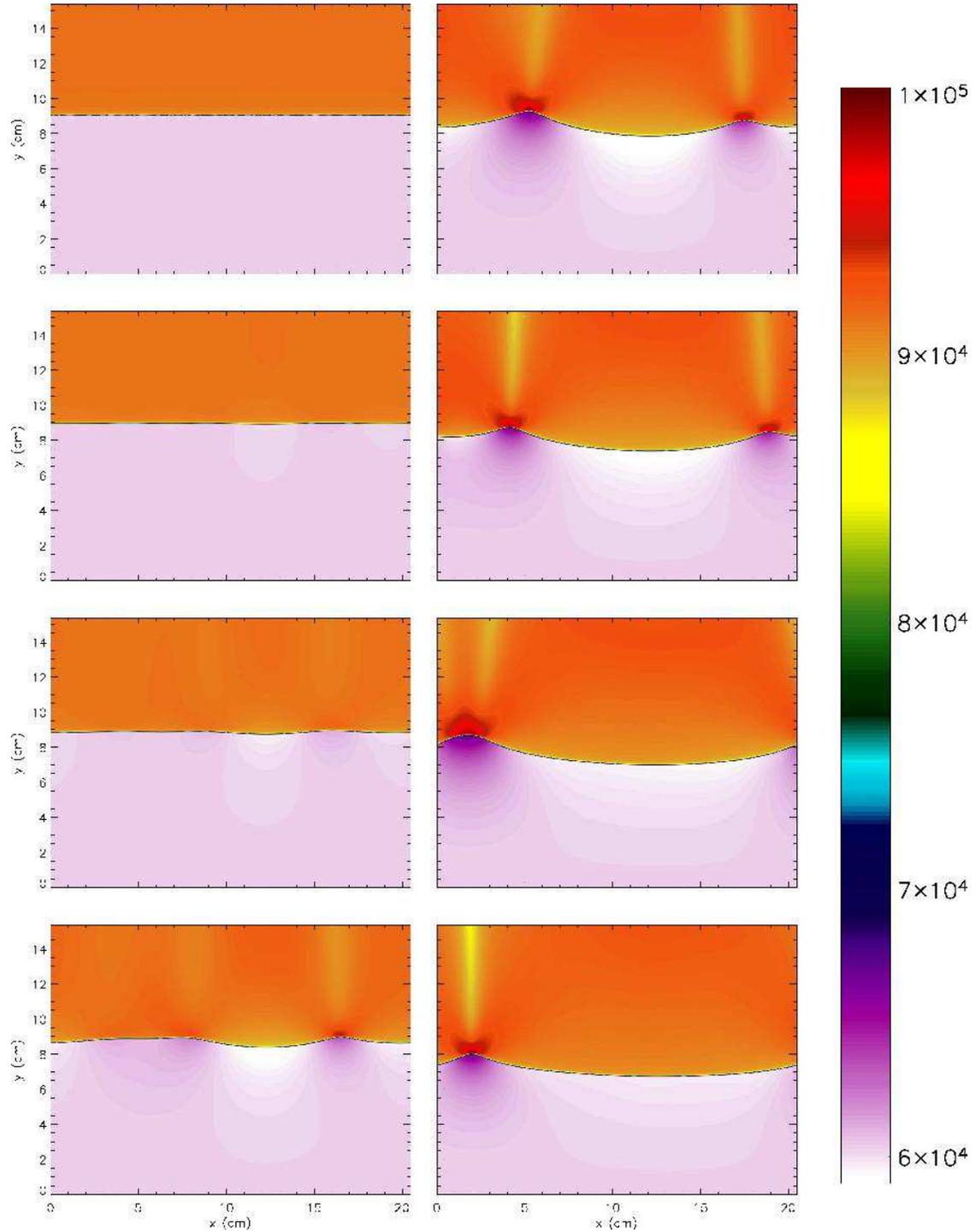}
\epsscale{1.0}
\end{center}
\caption{\label{fig:ld_4.e7_multimode} y-velocity for a $4\times
10^7~\gcc$ C/O flame undergoing a Landau-Darrieus instability.  In this
simulation, 30 different modes were perturbed.  Each pane is $3\times
10^{-4}$ s apart, spanning 0~s to $2.1\times 10^{-3}$~s.  The
initial perturbation rapidly evolves to only a few growing modes, and
quickly, two cusps form that dominate the flow.  Slowly, these two
cusps move toward each other and merge through the periodic
boundary, leaving a single, well-defined cusp behind.}
\end{figure}

As a motivation for the single mode studies, we look at a single case
of a multi-mode perturbation of a $4\times 10^7~\gcc$ flame in a
20.48~cm wide domain.  The flame front was perturbed by shifting the
zero point of the initial steady state flame by a sinusoid, with 30
different frequencies, whose phases and amplitudes were chosen randomly.  The maximum
amplitude of the perturbation is one flame thickness.
Figure~\ref{fig:ld_4.e7_multimode} shows the $y$-velocity at several
instances in time.  We see that immediately, the higher frequency
perturbations are washed out, and only the long wavelength modes
slowly begin to grow.  Several cusps begin to form, but quickly merge,
leaving two dominant cusps behind.  These remaining cusps gradually
move toward each other (through the periodic boundary), eventually
merging, leaving a single dominant cusp behind.  The timescale for the
cusps to merge is long, $\sim 2$~ms, so we show only one case here.
\citet{SNeCodePaper} shows another example, for a pure carbon flame at
a higher density, following the evolution for several cusp mergers.

\begin{figure}
\begin{center}
\plotone{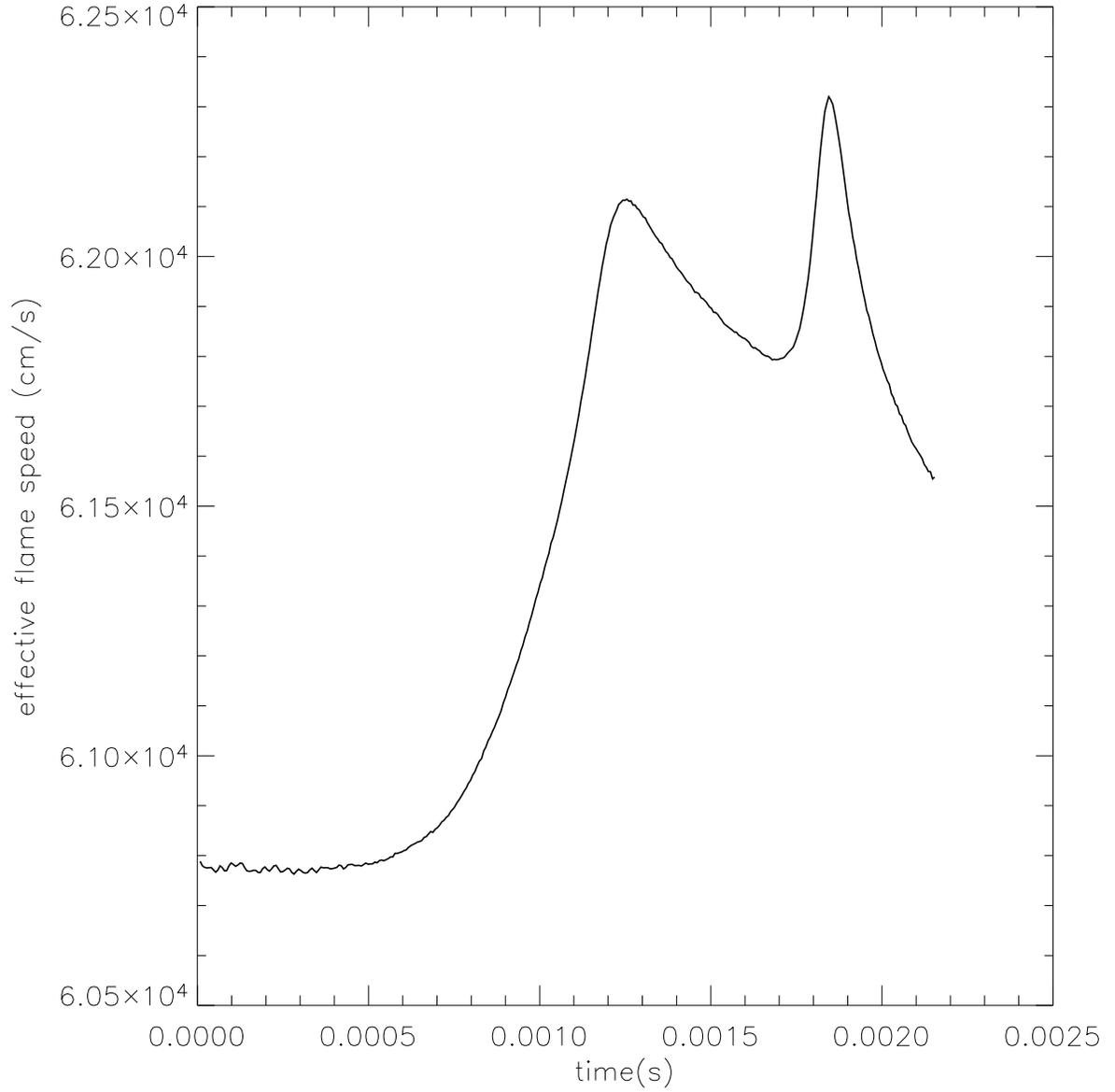}
\end{center}
\caption{\label{fig:ld_4.e7_multimode_speeds} Flame speed as a
function of time for the $4\times 10^7~\gcc$ C/O multimode
Landau-Darrieus instability, shown in Figure~\ref{fig:ld_4.e7_multimode}.}
\end{figure}

The flame speed can be measured by computing the carbon destruction
rate on the grid, taking into account the flow of material through
the boundaries:
\begin{equation}
F(t) = -\frac{\int_0^t\int_\Omega \rho \dot\omega_c d{\bf x}dt}
             {W (\rho X_c)^\mathrm{in}}
\end{equation}
where $\Omega$ is the spatial domain of the burning region, $W$ is the
width of inflow face, $(\rho X_c)^\mathrm{in}$ is the inflow carbon
mass fraction, and $\rho\dot\omega_c$ is the rate of consumption of
carbon due to nuclear burning.  The rate of change of this quantity in
a time interval $[T_1,T_2]$ is a measure of effective flame velocity:
\begin{equation}
V_\mathrm{eff} = \frac{\int_\Omega(\rho X_c)|^{T_2}_{T_1}d{\bf x}}{(T_2-T_1) W (\rho X_c)^\mathrm{in}}-u^\mathrm{in}
\label{EQN:veff}
\end{equation}
Figure~\ref{fig:ld_4.e7_multimode_speeds} shows the flame speed as a
function of time for the multimode calculation.  After an initial,
relatively constant velocity period, in which the initial
perturbations die out, the flame begins to accelerate rapidly.  The
velocity peaks shortly after the leftmost two initial cusps merge,
leaving behind two well defined cusps.  A second peak occurs at the
merging of these two cusps.

The merging of multiple cusps into a single dominant one has been
observed in other planar flame simulations (see for example
\citealt{gutman1990,helenbrooklaw1999}), and it seems to be a general
feature of the LD instability, with the exception of spherically
expanding flames \citep{blinnikovsasorov}.  We note however that we
see no evidence for a superposed cellular structure on the remaining
cusp, but this may be due to domain size restrictions
\citep{gutman1990}.  The final cusp that emerges is robust and shows
no signs of breaking down.

\subsection{Single-mode perturbations}

The main focus of the present study is to look at the acceleration
that a flame undergoing the LD instability will experience, in the
absence of any external forcing.  In particular, we wish to determine
whether there is any breakdown of the cusp configuration.  Since the
general consensus is that multimode perturbations in planar domains
tend to merge into a single, well defined cusp, we can seed a single
mode to study the behavior of this steady state configuration.

Table~\ref{table:singlemodeparams} list the parameters used for the
single-mode calculations.  In addition to a run of densities, we
considered the effect of the box width and resolution on the
instability for one choice of density ($4\times 10^7~\gcc$).  The
reference calculation at this density is a 10.24~cm wide $\times$
20.48~cm high domain, with a resolution of 6~points in the flame
width, $l_f$.  Numerical experiments have shown that this resolution
is enough to accurately follow the flame front (see
\citealt{SNeCodePaper}).  The critical wavelength for the growth of
1/2~carbon, 1/2~oxygen astrophysical flames over a range of densities
was computed in \citet{flame-curvature}, who, when converted to the
flame thickness definition used here, found $\sim 80$ flame
thicknesses to be the small scale cutoff.  If this result holds for
the densities we consider here, then for domains narrower than 80
flame widths, we do not expect the LD instability to be able to grow.
We note that our reference domain width is $\sim 174$ flame widths
wide, which should be well above the small scale cutoff for the LD
growth.  We will find this small scale cutoff directly by varying our
domain width, computing the Markstein number, and finding the zeros of
the LD dispersion relation, Equation~(\ref{eq:ld}).  A single sine
wave perturbation was used as an offset when mapping the steady state
laminar flame profile onto the domain.

\begin{figure}
\begin{center}
\epsscale{1.0}
\plotone{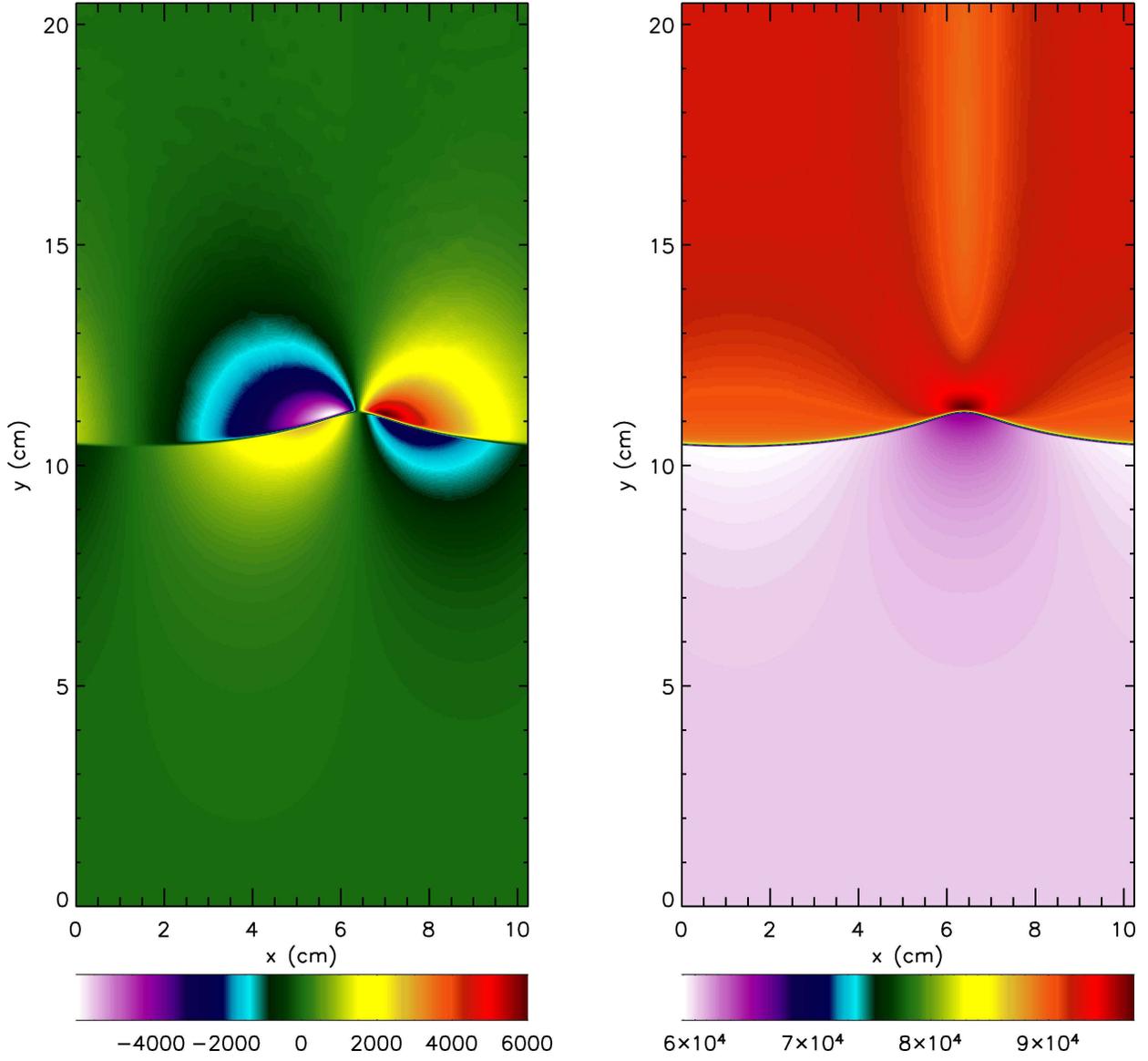}
\epsscale{1.0}
\end{center}
\caption{\label{fig:ld_4.e7} $x$-velocity (left) and $y$-velocity
         (right) for the high-resolution, 10.24~cm wide $4\times
         10^7~\gcc$ flame perturbed with a single mode after $5\times
         10^{-4}$~s.}
\end{figure}

\begin{figure}
\begin{center}
\epsscale{0.75}
\plotone{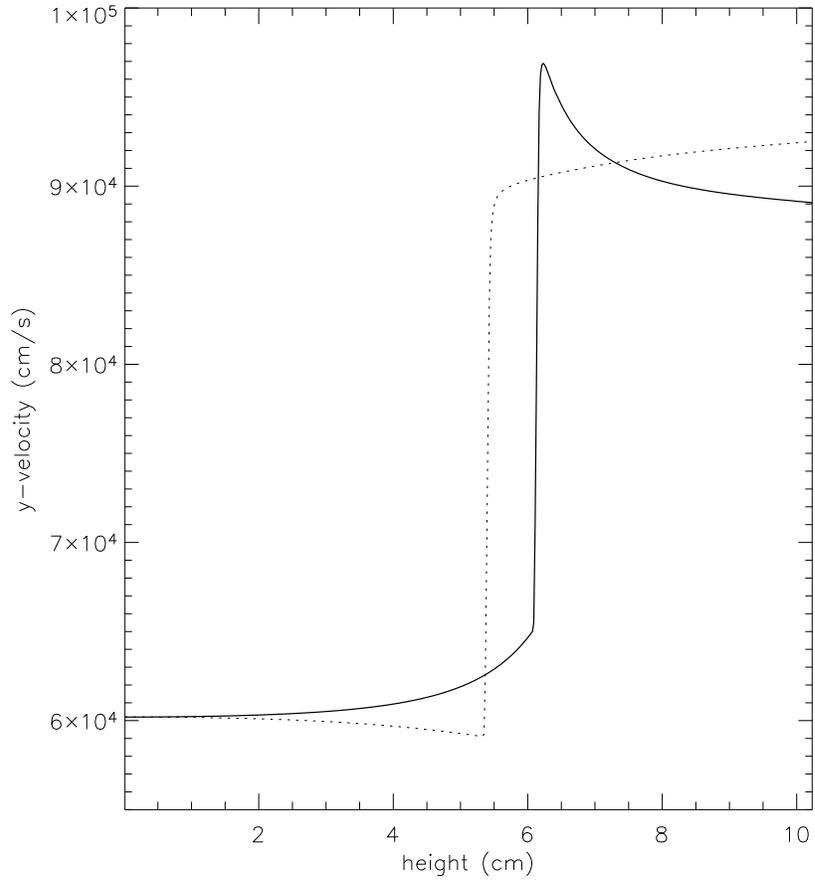}
\epsscale{1.0}
\end{center}
\caption{\label{fig:ld_4.e7_vel} $y$-velocity for the 10.24~cm wide,
         $4\times 10^7~\gcc$ flame perturbed when the flame velocity
         is maximum ($6.29\times 10^{-4}$ s) at the peak of the cusp
         (solid line) and the middle of the trough (dotted line).
         This demonstrates that the curvature leads
         to a change in the flame speed at different points along the
         flame front.}
\end{figure}

\begin{figure}
\begin{center}
\epsscale{0.75}
\plotone{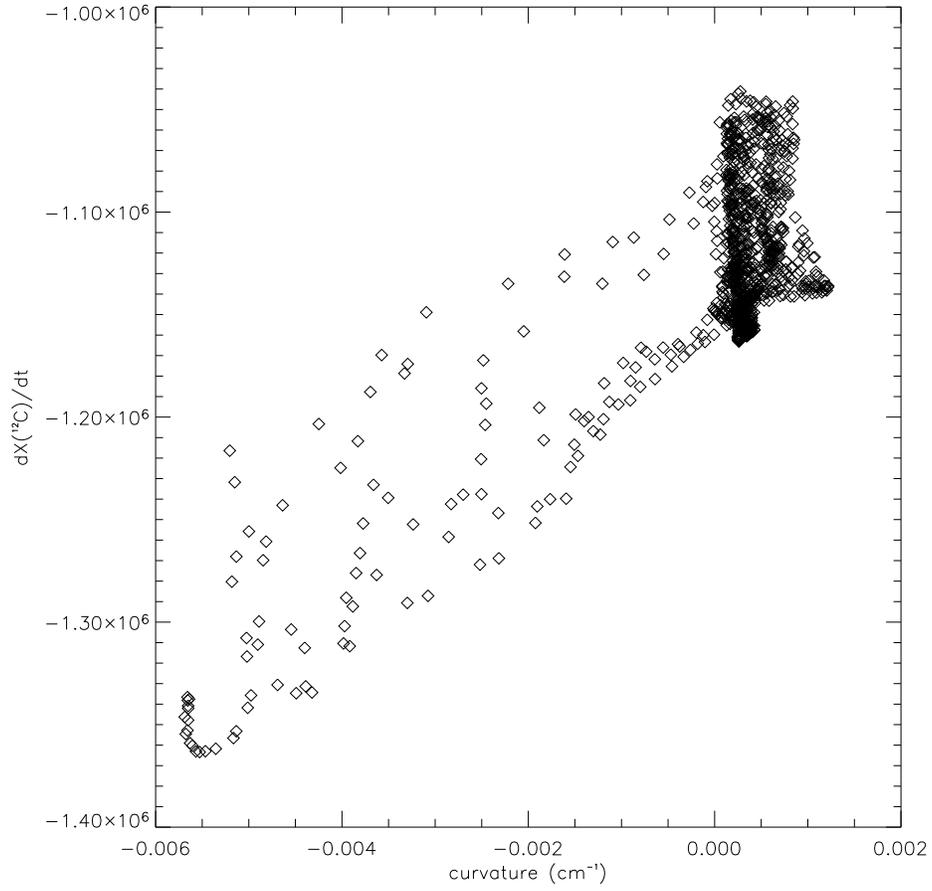}
\epsscale{1.0}
\end{center}
\caption{\label{fig:curvature} Carbon destruction rate vs.\ curvature
($d^2y_{\mathrm{interface}}/dx^2$) for the 10.24~cm wide, $4\times
10^7~\gcc$ flame.  When there is zero curvature, the carbon
destruction rate is the laminar value.}
\end{figure}

Figure~\ref{fig:ld_4.e7} shows the $x$- and $y$- velocities for the
reference $4\times 10^7~\gcc$ LD flame once the cusp has become well
defined.  After this point, the flame front retains this shape.  As
with the cusps in the multimode simulation, we see that the vertical
velocity peaks strongly right behind the cusp, and then drops below
the ambient post flame velocity as one moves further away from the
flame front.  One thing to note from these velocity figures is how
smooth the flow is ahead of and behind the flame.  There is virtually
no noise to disturb the pure LD instability.
Figure~\ref{fig:ld_4.e7_vel} illustrates this by showing two slices of
the $y$-velocity---through the center of the cusp and the center of
the trough---when the cusp is at its maximum.  This behavior is a
natural consequence of the local curvature and its effect on the
velocity streamlines.  The change in the peak carbon destruction rate
as a function of curvature is shown in Figure~\ref{fig:curvature}.
The curvature was computed by finding the position of the front and
taking the second derivative.  We see that the larger the magnitude of
the curvature, the greater the carbon destruction rate.  At zero
curvature, the carbon destruction rate is the laminar value.  The
scatter in the plot reflects the difficultly in defining the curvature
for a discretely defined interface.

\begin{figure}
\begin{center}
\epsscale{0.6}
\plotone{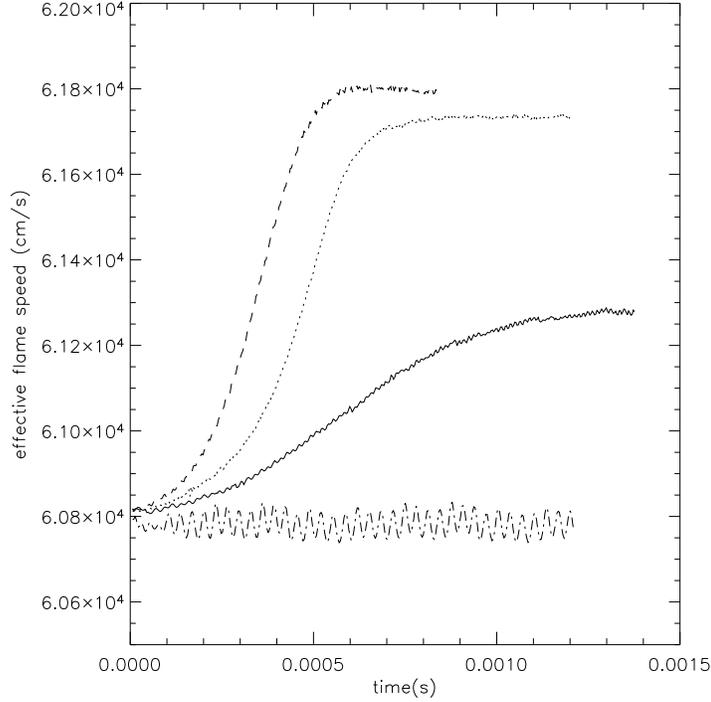}
\epsscale{1.0}
\end{center}
\caption{\label{fig:4.e7_flame_speeds} Flame speed vs.\ time for the
$4\times 10^7~\gcc$ LD unstable flame in a 20.48~cm domain (dot)
10.24~cm domain (dash), 5.12~cm domain (solid), and 2.56~cm domain
(dot-dash).  The slow growth of the 5.12~cm wide run reflects the
influence of this width being near the small scale cutoff for the
growth of the LD instability.  The 2.56~cm run shows no signs of
growing, as expected, since that domain width is well below the small
scale cutoff.  The 10.24~cm wide domain rises the fastest, as
expected, since this is closest to the maximum $\omega(k)$ of
Equation~[\ref{eq:ld}] for this flame.  }
\end{figure}

Figure~\ref{fig:4.e7_flame_speeds} shows the flame speed as a function
of time for the $4\times 10^7~\gcc$ C/O flame in several different
sized domains (2.56~cm, 5.12~cm, 10.24~cm, and 20.48~cm).  The flame
speed was computed by looking at the carbon consumption rate as in
Equation~(\ref{EQN:veff}).  As the
box width was varied, we choose to make $\delta y/W$ constant, where
$\delta y$ is the perturbation amplitude, and W is the box width.  All
other parameters (i.e., box height, boundary conditions, etc.) were
held constant across the different simulations.  The velocity of the
flame for the two widest domains increases quickly and reaches a peak,
where it levels off, at a value about 1.6\%~higher than the laminar
speed.  The slight decrease in the flame speed at late times for the
10.24~cm run is likely a result of the finite vertical extent of the
domain.  The 5.12~cm domain run evolves much more slowly.  This domain is
$\sim 87$ flame widths wide, right at the small scale cutoff for the
LD instability.  We believe that the slow growth that we see is a
manifestation of this cutoff.  The narrowest domain (2.56~cm) does not
grow at all.  It is not known what accounts for the small difference
in the asymptotic flame speeds for the two widest domains.

\begin{figure}
\begin{center}
\plottwo{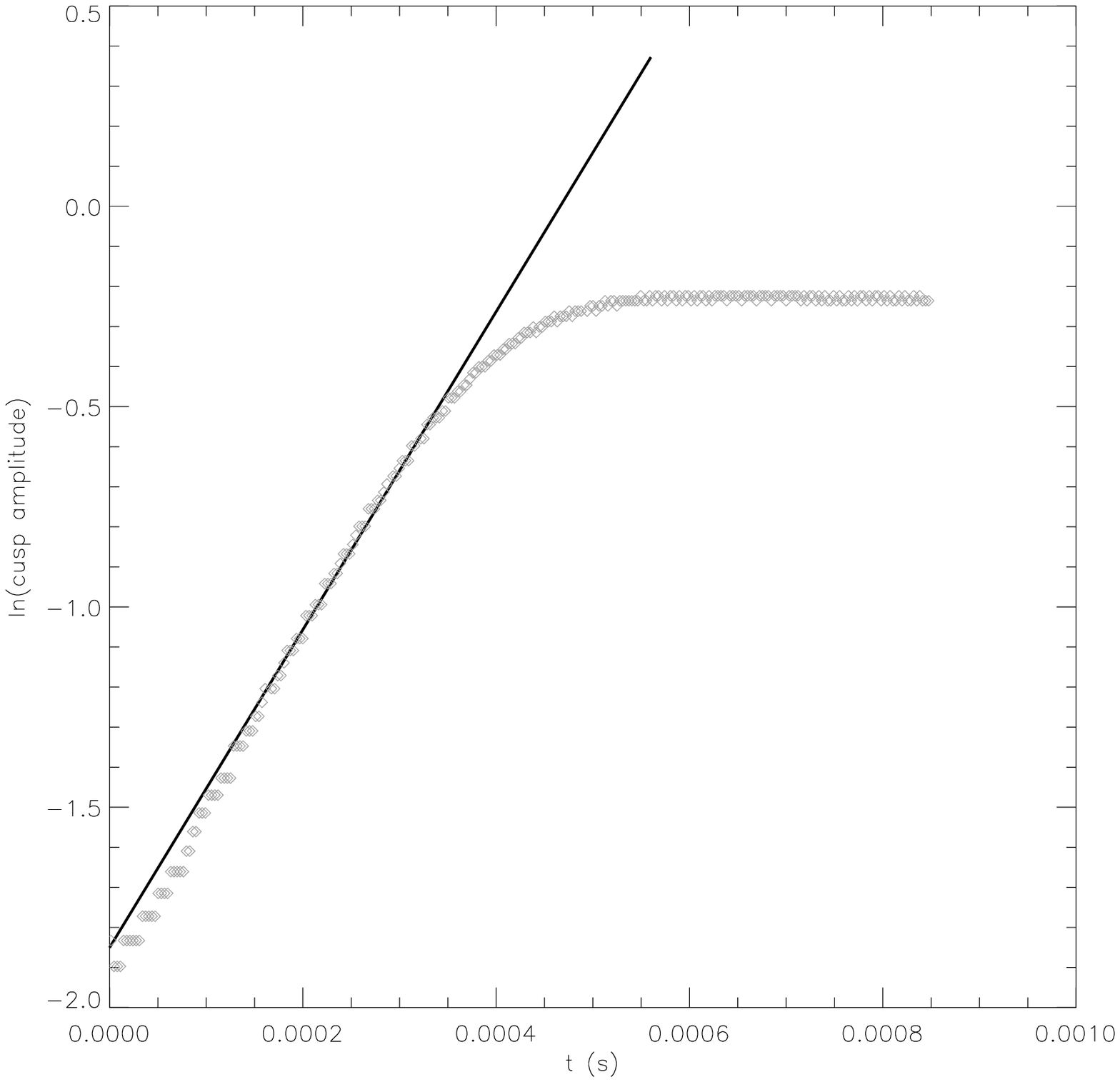}{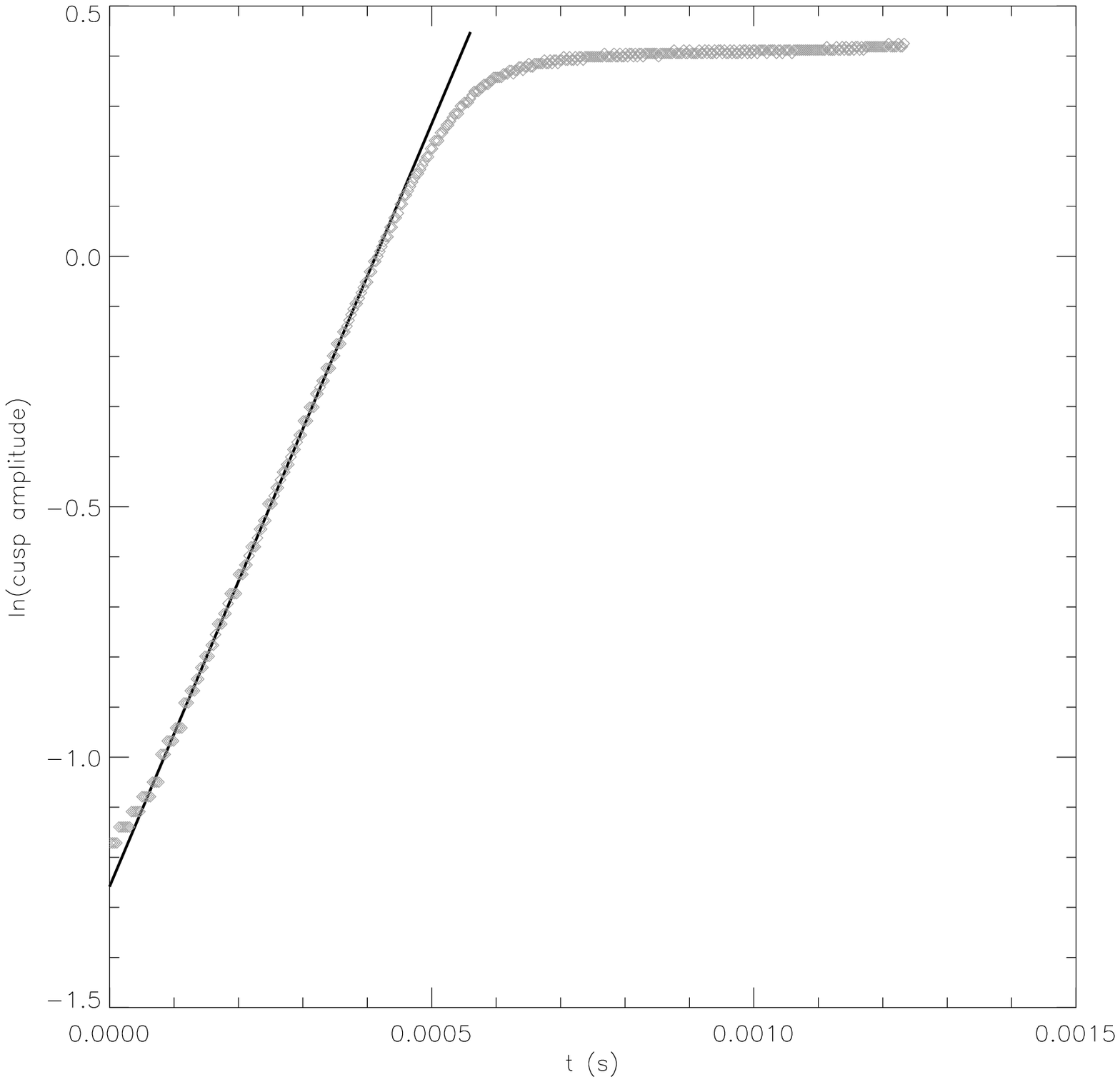}
\end{center}
\caption{\label{fig:ld_4.e7_cusp} Height of the cusp (natural log
scale) as a function of time (symbols) and a fit on an exponential to
the linear part (solid line) for the 10.24~cm wide domain (left) and
the 20.48~cm wide domain.}
\end{figure}

We can compare the time it takes for the flame speed to saturate to
the theoretical prediction of the LD growth rate,
Equation~(\ref{eq:ld}).  The amplitude of the cusp is computed by
laterally averaging the carbon mass fraction, and finding the
positions where it exceeds~0.05 and~0.45.
Figure~\ref{fig:ld_4.e7_cusp} shows the height of the cusp as a
function of time for the 10.24~cm and 20.48~cm wide runs.  Looking at
these plots, there is clearly a regime of linear growth followed by a
transition to the nonlinear regime where the cusping halts the growth,
and the amplitude reaches a steady state value.  The solid line is a
fit to
\begin{equation}
y = A e^{\omega t} \enskip ,
\end{equation}
for the interval of $1\times 10^{-4}$ to $4\times 10^{-4}$~s (the
linear regime), which gives $\omega = 3968$ and $3046$
respectively---the narrower domain grows faster.  We ignored the very
first part of the growth to avoid any transients.  If we were to
ignore the effects of curvature and the flame thickness (i.e., set
$\Ma = 0$ in Equation~[\ref{eq:ld}]), we would get predictions of $\omega =
8200$ and $4100$ for these domains.  Thus, the finite thickness of the
flame slows down the growth of the LD instability, demonstrating the
effects of the curvature on the flame speed. 

\begin{figure}
\begin{center}
\plotone{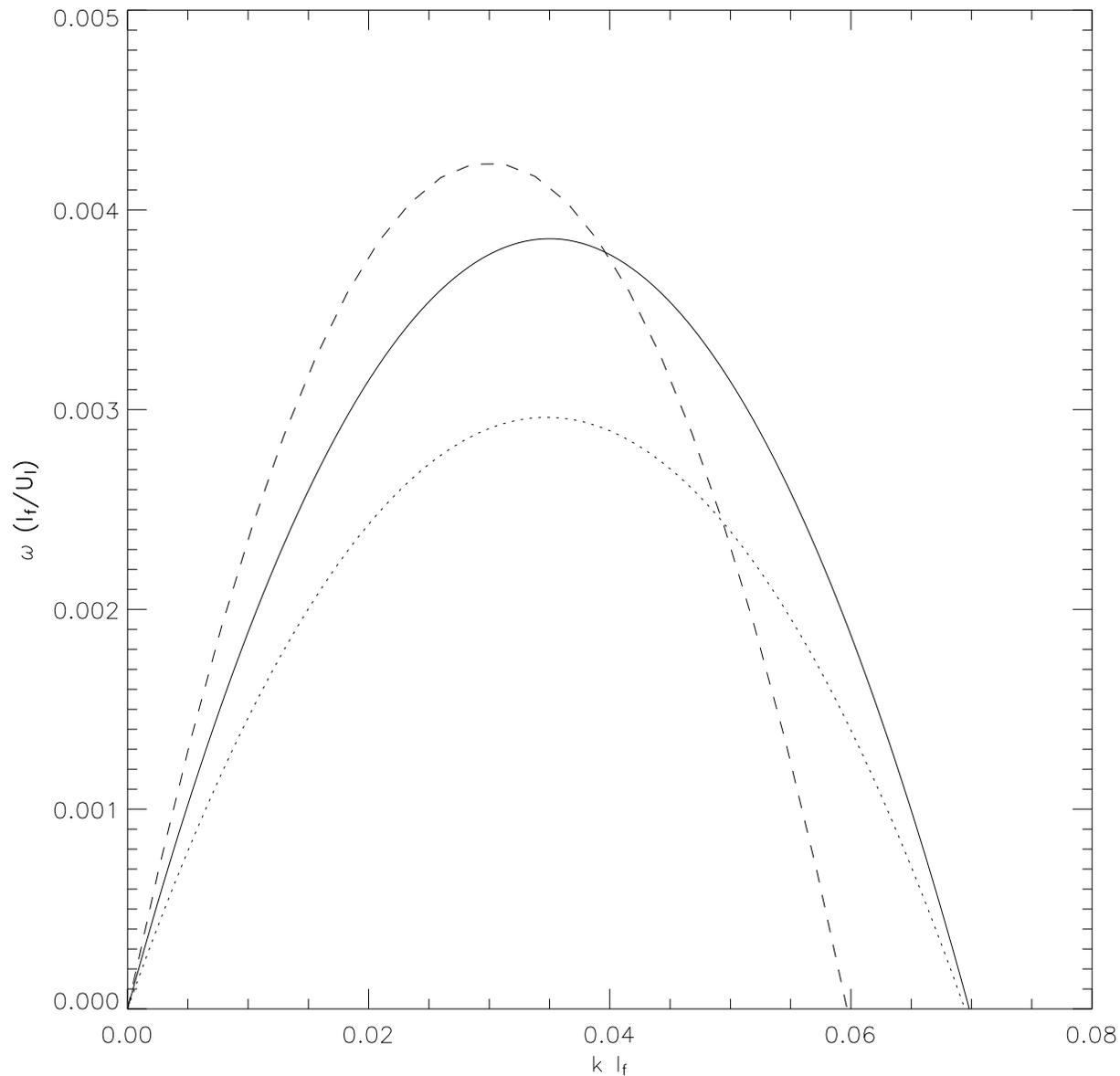}
\end{center}
\caption{\label{fig:dispersion} Predicted non-dimensional growth rate ($\omega
\; l_f/U_l$) of the LD instability as a function of dimensionless
wavenumber ($k \; l_f$) from Equation~[\ref{eq:ld}] for the three different
densities, $2\times 10^7~\gcc$ (dashed), $4\times 10^7~\gcc$ (solid),
and $8\times 10^7~\gcc$ (dotted), based on the measured Markstein numbers.}
\end{figure}

The finite thickness imposes a small scale cutoff to the LD growth,
which we can now compute.  Using the flame data from
Table~\ref{table:flameproperties}, we can solve for the Markstein
number, and find $\Ma = -2.45$ for both of these domain widths.  A
negative value is to be expected, as it indicates that a positively
curved flame burns more slowly, which is expected for our flames where
thermal diffusion dominates mass diffusion \citep{flame-curvature}.
With this value of the Markstein number, we can plot the growth rate
as a function of wavenumber, Figure~\ref{fig:dispersion}.  We see that
the LD stops growing for dimensionless wavenumbers larger than 0.07,
or domain widths smaller than 5.28~cm.  This is consistent with
behavior of the 2.56~cm and 5.12~cm runs we showed above.  For this
reason, we do not perform fits to these narrowest domain runs.  The
maximum growth should be seen for dimensionless wavenumbers of 0.035,
or a domain with of 10.6~cm.  Looking back at
Figure~\ref{fig:4.e7_flame_speeds}, we see that the 10.24~cm run (our
closest domain width to this maximum) reached the peak velocity
quickest and had the largest growth rate (as determined by the fits to
the cusp amplitude growth), consistent with the dispersion relation.
This also explains why the multimode perturbation presented above went
through an intermediate phase consisting of two cusps.  There, the
domain was 20.48~cm wide, and perturbations were made on many length
scales.  The dispersion relation predicts that the 10.6~cm
perturbations will grow fastest, overwhelming any smaller
perturbations.  This means that the mode that we seeded that put two
wavelengths in the box would grow the fastest, resulting in the
intermediate, moderately long lived state with two cusps.

\begin{figure}
\begin{center}
\epsscale{0.6}
\plotone{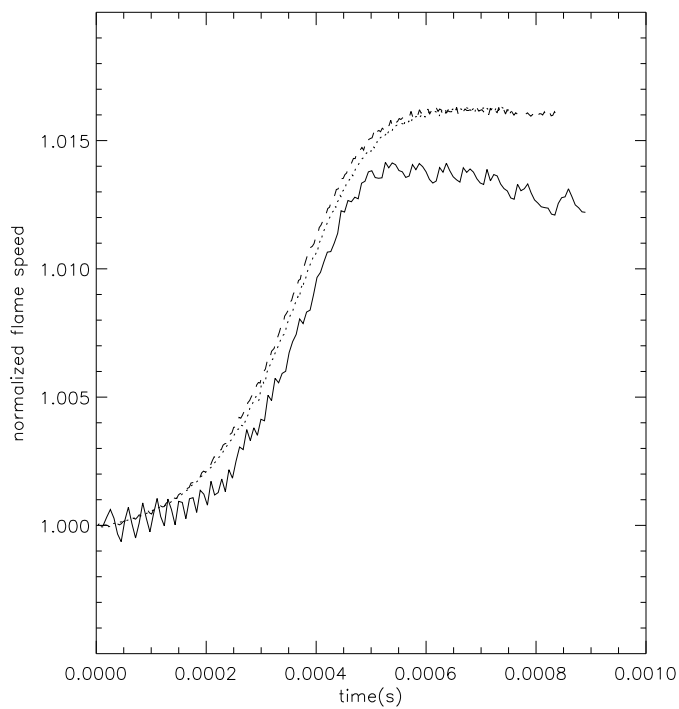}
\end{center}
\caption{\label{fig:4.e7_res_study} Normalized flame speed,
$V_{\mathrm{eff}}(t)/V_{\mathrm{eff}}(t=0)$, vs.\ time for the $4\times 10^7~\gcc$ LD unstable in a
10.24~cm wide domain at 3 different resolutions: 3.0~zones/$l_f$
(solid), 5.9~zones/$l_f$ (dash), and 11.8~zones/$l_f$ (dot).  The two
highest resolution runs seem to have converged.  The main calculations
presented in this paper were computed at the 5.9~zones/$l_f$
resolution.  The normalization is required to correct for the small
differences in the laminar flame speed with resolution.}
\end{figure}

Figure~\ref{fig:4.e7_res_study} shows the flame speed vs.\ time for
the 10.24~cm wide, $4\times 10^7~\gcc$ run at three different resolutions.
The flame speed is normalized to the laminar speed at that resolution
to account for the small resolution dependence on this speed.
Table~\ref{table:res_study} lists the laminar flame speeds for the
different resolutions.  At the highest resolutions, the laminar flame
speeds differ by 0.13\%, and the normalized velocity vs.\ time curves
lie on top of one another, demonstrating convergence.  The coarsest
resolution laminar flame speed differs by 1.1\%, and the evolution
with time is clearly not converged.  All the simulations presented in
this paper are performed at this middle resolution or higher.  This
resolution agrees with that found in the corresponding resolution
study in \citet{SNeCodePaper}.

\begin{figure}
\begin{center}
\plotone{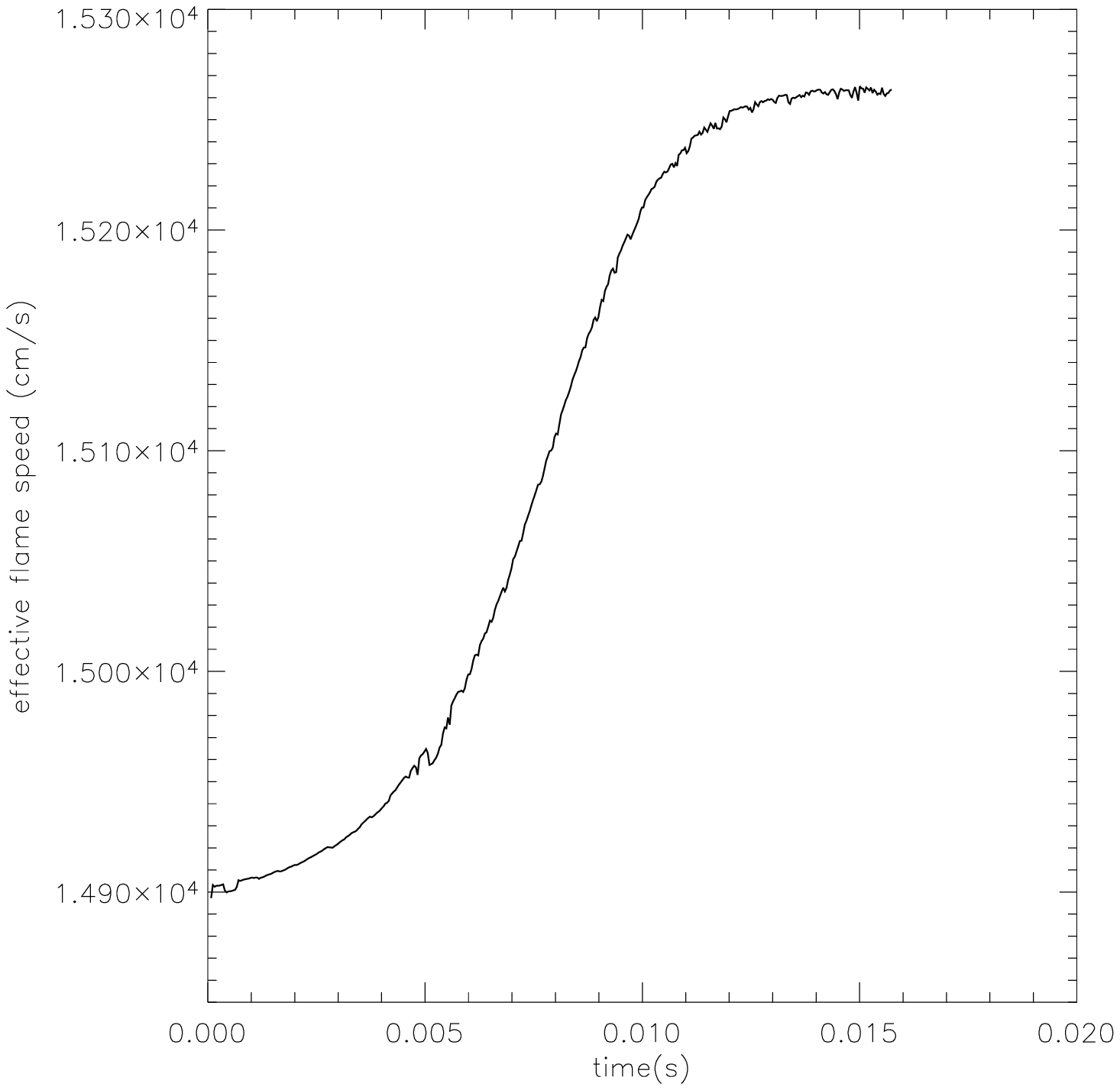}
\end{center}
\caption{\label{fig:2.e7_flame_speed} Flame speed as a function of
  time for the $2\times 10^7~\gcc$ LD unstable flame.}
\end{figure}

\begin{figure}
\begin{center}
\plotone{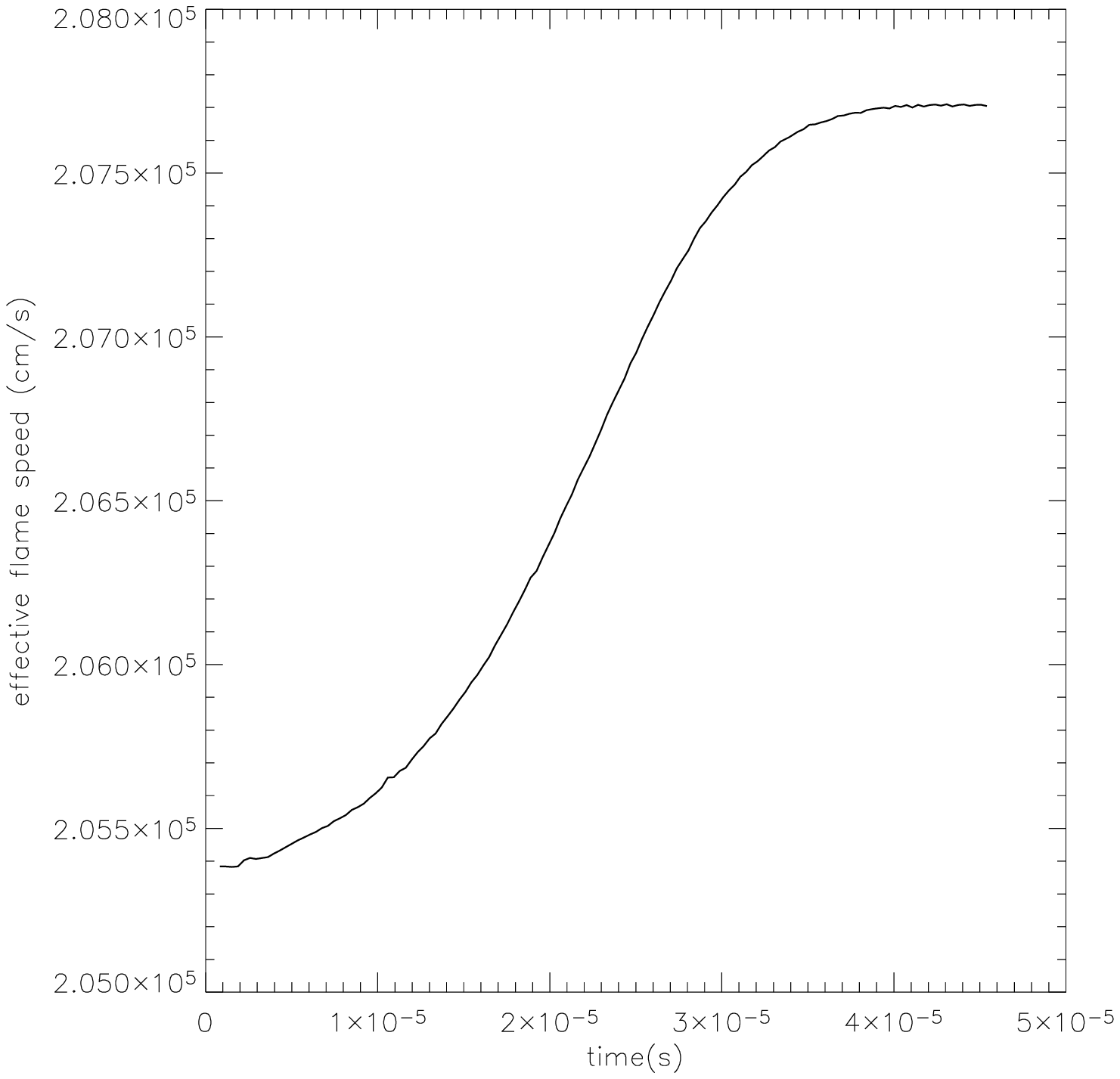}
\end{center}
\caption{\label{fig:8.e7_flame_speed} Flame speed as a function of
  time for the $8\times 10^7~\gcc$ LD unstable flame.}
\end{figure}

\begin{figure}
\begin{center}
\plotone{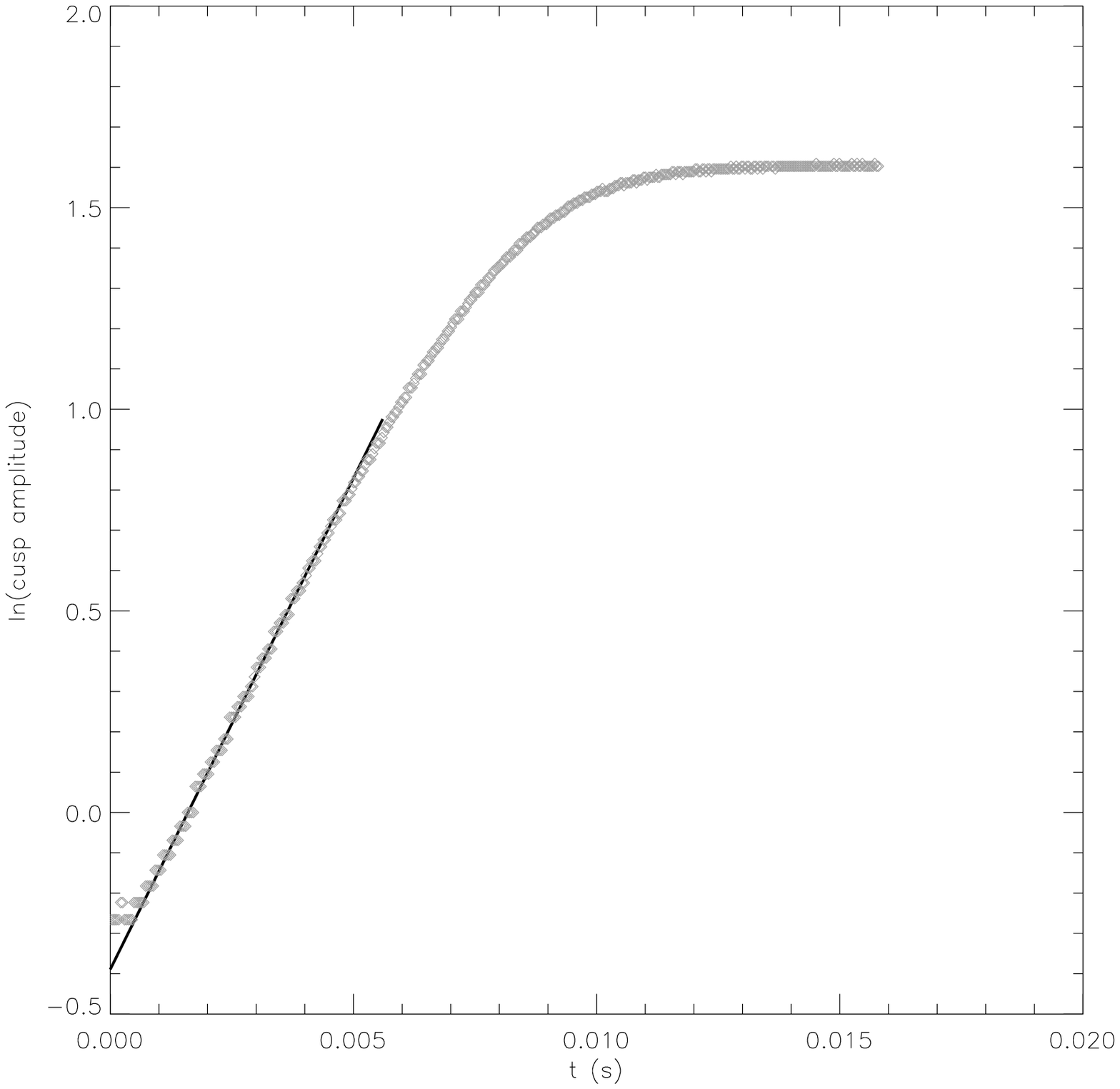}
\end{center}
\caption{\label{fig:2.e7_cusp} Cusp amplitude (natural log scale) as a
function of time (symbols) and a fit on an exponential to the linear
part (solid line) for the $2\times 10^7~\gcc$ flame, in the interval
[$10^{-3}$~s, $4\times 10^{-3}$~s].  The growth rate of the fit is
$\omega = 244~\mathrm{s}^{-1}$.}
\end{figure}

\begin{figure}
\begin{center}
\plotone{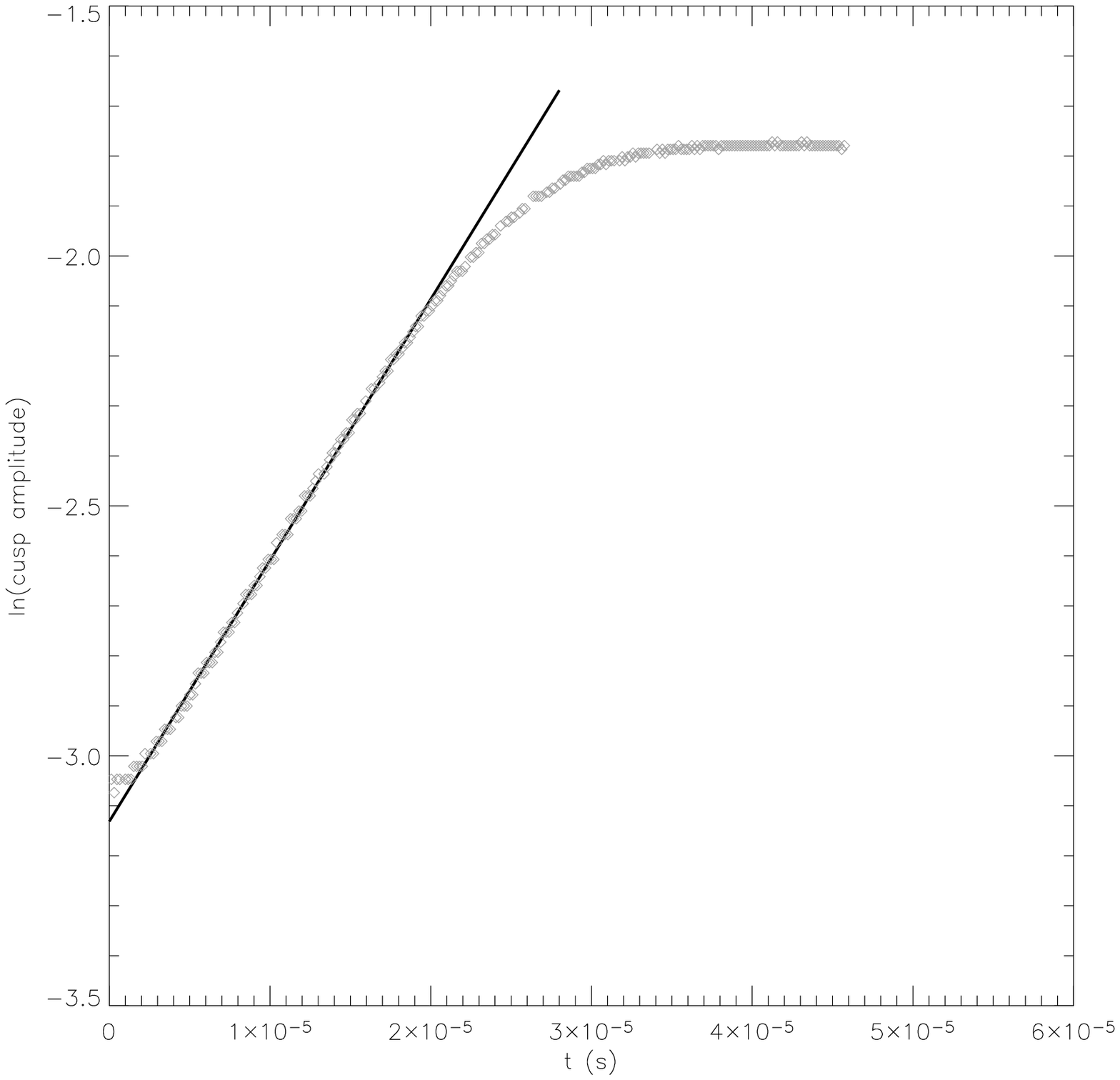}
\end{center}
\caption{\label{fig:8.e7_cusp} Cusp amplitude (natural log scale) as a
function of time (symbols) and a fit on an exponential to the linear
part (solid line) for the $8\times 10^7~\gcc$ flame, in the interval
[$5\times 10^{-6}$~s, $2\times 10^{-5}$~s].  The growth rate of the
fit is $\omega = 522000~\mathrm{s}^{-1}$.}
\end{figure}

Figures~\ref{fig:2.e7_flame_speed} and \ref{fig:8.e7_flame_speed} show
the flame speed as a function of time for the $2\times 10^7~\gcc$ and
$8\times 10^7~\gcc$ runs respectively.  The evolution at these
densities proceeded in much the same way as the $4\times 10^7~\gcc$
run shown in Figure~\ref{fig:ld_4.e7}, so we do not show the velocity
fields.  Only one domain width was used for these calculations, with
the size picked close to the wavelength that maximizes the growth
rate.  The change in density (and expansion parameter across the
flame) changes the speedup we get.  The $2\times 10^7~\gcc$ flame
accelerates by 2.4\% and the $8\times 10^7~\gcc$ flame by 1.1\%.
Together with the results for the $4\times 10^7~\gcc$ flame, we see
that the acceleration is larger for the lower density\slash larger expansion
ratio flames.  We can compute the Markstein numbers for these flames
in the same manner as above and find $-3.42$ and $-2.02$ for the
$2\times 10^7~\gcc$ and $8\times 10^7~\gcc$ flames respectively (see
Figures~\ref{fig:2.e7_cusp} and~\ref{fig:8.e7_cusp}).  Again, the fit
of an exponential to the linear portion of the growth is excellent.
The dispersion relations for these flames are also plotted in
Figure~\ref{fig:dispersion}.  To fit them all on the same plot, the
curves are scaled by the flame speed and width.  The magnitude of the
Markstein number increases with decreasing density and increasing
expansion parameter.  This is not unexpected, and confirms the trend
seen in \citet{flame-curvature} for similar flames.  This suggests
that a lower density flame is harder to bend.

\begin{figure}
\begin{center}
\epsscale{1.0}
\plotone{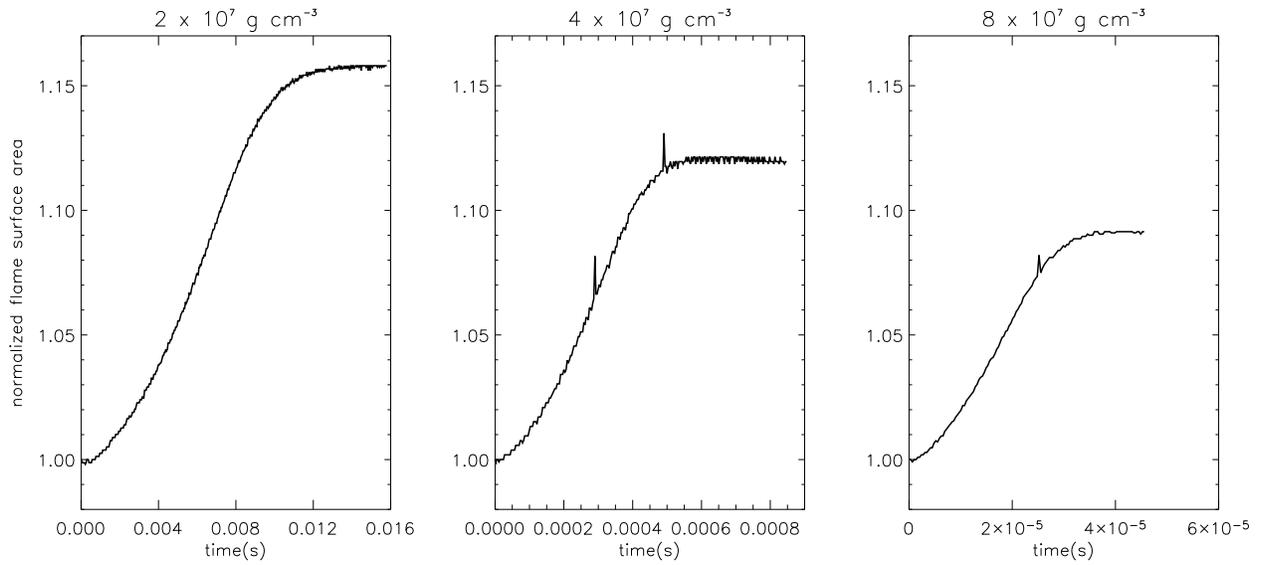}
\end{center}
\caption{\label{fig:flamelength} Normalized flame length,
$L(t)/L(t=0)$, as a function of time for the three different
densities, $2\times 10^7~\gcc$ (left), $4\times 10^7~\gcc$ (center),
and $8\times 10^7~\gcc$ (right).  The increase in flame length is
several times larger than the corresponding increase in the flame
speed.}
\end{figure}

As a final measure of the effects of curvature on the flame speed, we
can compare the growth in the flame surface area to the growth in
flame speed.  In the simplest model, these two are directly related in
a purely geometrical fashion, and the flame speed would be simply
\begin{equation}
\label{eq:speed_area}
v(t) = v_0 \frac{A(t)}{A_0} \enskip .
\end{equation}
We can measure the length of the flame surface by counting the number
of zone edges where the carbon mass fraction passes through 0.25.
This measure agrees well with computing the length of a contour as
computed with the commercial {\tt IDL} package, after normalization.
Since the flame surface is always sharp, this is a well-defined
procedure.  Figure~\ref{fig:flamelength} shows the normalized flame
length as a function of time for the three different densities (the
10.24~cm reference run was used for the $4\times 10^7~\gcc$ case).  In
all these cases, the increase in the flame length is several times
larger than the increase in the flame speed, showing a strong
deviation from the linear scaling of Equation~(\ref{eq:speed_area})---for
example, the $4\times 10^7~\gcc$ flame increases in speed by about
1.6\%, but the area increase is 12\%.  This deviation is due to
the effects of the localized curvature.  We note that this deviation
is larger in magnitude than that found in the compressible,
flame-model calculations of \citet{rhn2003}.

% ============================================================================
%  Conclusions
% ============================================================================
\section{CONCLUSIONS}
\label{sec:conclusions}

We presented direct numerical simulations of the Landau-Darrieus
instability at low densities in conditions relevant for Type Ia
supernova explosions.  We employed a low Mach number hydrodynamics
method---new in the application to astrophysics---to advance the flows
and enable us to follow the development of the LD instability from the
linear regime through the nonlinear regime.  Resolving the flame
ensures that we account for the full coupling between the flame and
the flow and the effects of curvature.  The consistency of our results
with the theoretical predictions validate this approximation for
astrophysical flames.  Fully compressible versions of the same
simulations would not be possible, or prohibitively expensive and
error prone.

Acceleration of the flame was seen, due to the increase of flame
surface, until the nonlinear cusp formation sets in, at which point
the velocity stabilized a few percent higher than the laminar speed.
The maximum increase in flame speed observed was $\sim 2\%$---this is
in contrast to the results presented in \cite{rhn2003}, where
accelerations of $\sim 30\%$ were observed.  There are several
possible reasons to explain these differences.  Most of the numerical
results for the LD instability treat the flame as infinitely thin.
Here we've included the effects of the finite thickness of the flame,
and therefore, implicitly included curvature effects.  At the
densities we consider, the burning is expected to terminate at the
intermediate mass elements, so we burn to magnesium whereas they burn
to nickel.  Furthermore we use half carbon/half oxygen fuel instead of
pure carbon.  This will affect the expansion factor across the flame.
Finally, fully understanding the role of the numerics requires a
detailed comparison of the different methods, using the same input
physics, and could potentially be a research project in the future.

The speedup increases with decreasing density (or increasing expansion
parameter), as expected.  With time, it seems as if the velocities
asymptote to the same value, for a given density.  When a multimode
perturbation is considered, it is observed that the cusps eventually
merge into a single cusp.

We were able to perform fits of the growth of the cusp amplitude with
time to compute the growth rates.  Using the theoretical prediction
(Equation~[\ref{eq:ld}]), we were able to compute the Markstein number for
these flames, finding values $< -2.0$, and growing in magnitude with
decreasing density.  These values and the trend with density are
consistent with the results of \citet{flame-curvature}.  For the
$4\times 10^7~\gcc$ density run, the large number of domain widths
allowed us to compare to the predictions of the fastest growing mode
and small scale cutoff from the theoretical dispersion relation, and
we found strong agreement.  Furthermore, this demonstration of the
small scale cutoff and the deviation from the infinitely thin flame
($\Ma = 0$) estimates of the growth rate, as well as the difference in
the growth of the flame speed and area emphasize the importance of
curvature in small-scale flame dynamics.  As we zoom out from these
fully resolved flames, several steps will be needed at intermediate
scales between the fully resolved flame, where curvature effects
appear important, and the full star, with validation of the flame
model on each scale through direct comparison to the immediately
smaller scale simulations.  The present direct numerical simulations
provide the anchor for this subgrid model validation.  These curvature
effects will be explored more fully in paper II, focusing on the
reactive Rayleigh-Taylor instability.

We saw no breakdown of the nonlinear stabilization mechanism as hinted
to in the calculations presented by \citet{niemeyerhillebrandt1995a}.
Furthermore, none of the flames we studied showed a superposed
cellular structure, although, there is some evidence
\citep{gutman1990} that this phenomena is present only on very large
scales, and \citet{blinnikovsasorov} admit that it may actually be
seeded by numerical noise.  The widest domain that we were able to
simulate was $\sim 350$ flame widths.  \citet{helenbrooklaw1999} only
saw this instability when they increased their domain size to 20~times
the wavelength where the LD growth rate is the maximum.  This would
translate into a domain size of $230$~cm for the $4\times 10^7~\gcc$
flame, which would require a 23,000 zone wide simulation if we are to
resolve the flame.  The growth rate for a domain that wide would be
very small, requiring a much longer simulation.  This is very
expensive, and furthermore, it is not known if this factor of 20 is
universal.  We are not aware of any resolved calculations of the LD
instability that have shown this secondary instability.

We note that all of our calculations considered planar geometries
only.  On the scales where the spherical geometry of the star is
likely to make a difference, the flame will be disturbed by the larger
scale motions in the star, other instabilities as well (notably the
Rayleigh-Taylor instability, presented in \citealt{SNrt}) and
subjected to the interaction with flame-generated turbulence.  All of
these processes will dominate the pure LD instability,
considered in the present paper.

\acknowledgements 

The authors thank L.~J. Dursi for useful comments on the manuscript
and F.~X. Timmes for making his equation of state and conductivity
routines available on-line.  Support for this work was provided by the
DOE grant number DE-FC02-01ER41176 to the Supernova Science
Center/UCSC and the Applied Mathematics Program of the DOE Office of
Mathematics, Information, and Computational Sciences under the
U.S. Department of Energy under contract No.\ DE-AC03-76SF00098.  The
calculations presented here were performed on the IBM Power4 (cheetah)
at ORNL, sponsored by the Mathematical, Information, and Computational
Sciences Division; Office of Advanced Scientific Computing Research;
U.S. DOE, under Contract No.\ DE-AC05-00OR22725 with UT-Battelle, LLC
and the UCSC UpsAnd cluster supported by an NSF MRI grant AST-0079757.

%\clearpage

% ============================================================================
%  Figures
% ============================================================================

\clearpage

\begin{table}
\begin{center}
\caption{\label{table:flameproperties} Flame properties for 0.5 $^{12}$C/0.5 $^{16}$O flames.}
\begin{tabular}{rdrd}
\tableline
\tableline
\multicolumn{1}{c}{$\rho$}  & \multicolumn{1}{c}{${\alpha}$\tablenotemark{a}} & \multicolumn{1}{c}{$v_{\mathrm{laminar}}$}  & \multicolumn{1}{c}{$l_f$} \\
\multicolumn{1}{c}{($\gcc$)} & & \multicolumn{1}{c}{($\cms$)} & \multicolumn{1}{c}{(cm)} \\
\tableline
$2\times 10^7$ & 1.69 & $1.50\times 10^4$ & 0.26 \\
$4\times 10^7$ & 1.52  & $6.08\times 10^4$ & 0.059 \\
$8\times 10^7$ & 1.39  & $2.04\times 10^5$ & 0.011 \\
\tableline
\end{tabular}
\end{center}
\tablenotetext{a}{$\alpha = \rho_{\mathrm{fuel}}/\rho_{\mathrm{ash}}$}
\end{table}

\begin{table}
\begin{center}
\caption{\label{table:singlemodeparams} Parameters for single mode
Landau-Darrieus simulations.}
\begin{tabular}{rlrrrr}
\tableline
\tableline
\multicolumn{1}{c}{$\rho$}& description & \multicolumn{1}{c}{$W \times H$} & \multicolumn{1}{c}{fine grid}  & $W/l_f$ & $l_f/\Delta x$ \\
\multicolumn{1}{c}{($\gcc$)} & & \multicolumn{1}{c}{(cm)} \\
\tableline
$2\times 10^7$ & \nodata  & 51.20 $\times$ 102.4\phn & 1536 $\times$ 3072 & 199.2 & 7.7 \\[2mm]

$4\times 10^7$ & very narrow     & 2.56  $\times$ \phn20.48 &  256 $\times$ 2048  & 43.4  & 5.9 \\
               & narrow          & 5.12  $\times$ \phn20.48 &  512 $\times$ 2048  & 86.8  & 5.9 \\
               & regular         & 10.24 $\times$ \phn20.48 & 1024 $\times$ 2048  & 173.6 & 5.9 \\
               & wide            & 20.48 $\times$ \phn20.48 & 2048 $\times$ 2048  & 347.2 & 5.9 \\
               & low-resolution  & 10.24 $\times$ \phn20.48 & 512 $\times$ 2048  & 173.6 & 3.0 \\
               & high-resolution & 10.24 $\times$ \phn20.48 & 2048 $\times$ 4096  & 173.6 & 11.8 \\[2mm]

$8\times 10^7$ & \nodata & 2.56 $\times$ \phn\phn3.84 & 2048 $\times$ 3072 & 232.7 & 8.8 \\
\tableline
\end{tabular}
\end{center}
\end{table}

\begin{table}
\begin{center}
\caption{\label{table:res_study} $4\times 10^7~\gcc$ laminar flame speed vs.~resolution}
\begin{tabular}{rr}
\tableline
\tableline
\multicolumn{1}{c}{$l_f/\Delta x$} & \multicolumn{1}{c}{$v_{\mathrm{laminar}}$} \\
               &  \multicolumn{1}{c}{($\mathrm{cm~s^{-1}}$)} \\
\tableline
3.0  & $6.013\times 10^4$ \\
5.9  & $6.074\times 10^4$ \\
11.8 & $6.082\times 10^4$ \\
\tableline
\end{tabular}
\end{center}
\end{table}

\clearpage

\clearpage

\clearpage

\clearpage

\clearpage

\clearpage

%\clearpage

%\begin{figure}
%\begin{center}
%\plotone{ld_8.e7.eps}
%\end{center}
%\caption{\label{fig:ld_8.e7} Results of the $8\times 10^7~\gcc$ LD
%flame simulation at $4.6\times 10^{-5}$ s.  Shown are the $x$-velocity
%(left) and $y$-velocity (right).  Aside from the length and velocity
%scales, these images look qualitatively the same as the $4\times
%10^7~\gcc$ results (Fig.~\ref{fig:ld_4.e7}).}
%\end{figure}

\clearpage

\clearpage

\end{document}